\RequirePackage{fix-cm}
\documentclass[twocolumn,epjc3]{svjour3}  
\smartqed  % flush right qed marks, e.g. at end of proof
\RequirePackage{graphicx}
\usepackage{amsmath, amsfonts,amssymb,amsxtra}
\journalname{Eur. Phys. J. C}
\begin{document}

\title{Gauge fields with respect to $d=(3+1)$ in the Kaluza-Klein
  theories  and in the {\it spin-charge-family theory}
 %Insert your title here%\thanksref{t1}
}
%\subtitle{Do you have a subtitle?\\ If so, write it here}

%\titlerunning{Short form of title}        % if too long for running head

\author{Dragan Lukman %\thanksref{e1,addr1}
        \and
        Norma Susana Manko\v c Bor\v stnik \thanksref{e2}%,addr2,addr3} %etc.
}

%\thankstext{t1}{Grants or other notes
%about the article that should go on the front page should be
%placed here. General acknowledgments should be placed at the end of the article.
\thankstext{e2}{e-mail: norma.mankoc@fmf.uni-lj.si}

%\authorrunning{Short form of author list} % if too long for running head

\institute{University of Ljubljana, FMF, Dept. of  Physics, Jadranska 19, 1000 Ljubljana,
%\\ 
Slovenia %\label{addr1}
          % \and
           %Second address \label{addr2}
           %\and
           %\emph{Present Address:} if needed\label{addr3}
}

\date{Received: date / Accepted: date}
% The correct dates will be entered by the editor

\maketitle

\begin{abstract}
It is shown that in the {\it spin-charge-family} theory~\cite{IARD2016,%
norma2014MatterAntimatter,JMP2015,JMP2013,n2012scalars}, as well as in all the
Kaluza-Klein like theories~\cite{mil,zelenaknjiga}, vielbeins and spin connections manifest in 
$d=(3+1)$ space equivalent vector gauge fields, when space with $d\ge5$ has large enough 
symmetry. 
The authors demonstrate this equivalence in spaces with the symmetry of the metric tensor in 
the space out of $d=(3+1)$ - $g^{\sigma \tau} = \eta^{\sigma \tau} \,f^{2}$ - for any scalar 
function $f$ of the coordinates $x^{\sigma}$, where  $x^{\sigma}$ denotes coordinates of 
space out of  $d=(3+1)$.
Also the connection between vielbeins and scalar gauge fields in $d=(3+1)$ (offering the
explanation for the Higgs's scalar)  is discussed. 
\keywords{Unifying theories\and Beyond the standard model \and Properties of scalar fields \and 
Origin and properties of gauge bosons \and  Kaluza-Klein-like theories}
 \PACS{12.60.-i  \and 11.10.Kk \and 04.50.-h \and 12.10.-g \and 11.30.-j  \and14.80.-j  \and
11.15.Ex  \and 12.90.+b}
% \subclass{MSC code1 \and MSC code2 \and more}
\end{abstract}
\section{Introduction } 
\label{introduction}

The {\it spin-charge-family} theory~\cite{IARD2016,norma2014MatterAntimatter,JMP2015,%
JMP2013,n2012scalars} explains, starting with the simple action~(Eq.~(\ref{wholeaction})) in 
$d>(3+1)$, all the assumptions of the {\it standard model}, as well as other phenomena, like
the matter-antimatter asymmetry, dark matter appearance and others. In this theory the spin
connection fields manifest in the low energy regime as the known vector gauge fields as well 
as the Higgs's scalars (and the Yukawa couplings), while in the Kaluza-Klein theories~\cite{mil,%
zelenaknjiga} vielbeins (or rather metric tensors) are usually used to represent vector gauge fields.   

We demonstrate in this paper that in $d$-dimensional spaces with the symmetry  of the metric 
tensor in~$(d-4)$-dimensional space $g_{\sigma \tau}$ $= \eta_{\sigma \tau}\, f^{-2}$ [where 
($x^{\sigma}$, $x^{\tau}$) determine the coordinates of the almost compactified 
space~\cite{hope,norma2004,hn2006}, $\eta_{\sigma \tau}$ is the diagonal matrix in this space and $f$ is any  
scalar function of these coordinates] both procedures - the ordinary Kaluza-Klein one with 
vielbeins and the one with spin connections (related to the vielbeins, Eq.(\ref{omegaabe1m})), 
used in the {\it spin-charge-family} theory~(\cite{IARD2016,norma2014MatterAntimatter,%
JMP2015,JMP2013,n2012scalars} and the references therein) - lead in $d=(3+1)$ to the same 
vector gauge fields. That either the vielbeins or the spin connections represent in symmetric 
enough $(d-4)$ spaces in $d=(3+1)$ the same vector gauge fields is known for a long 
time~\cite{mil,zelenaknjiga,norma2004}. 

This paper is to clarify the equivalence of representing in theories with higher dimensional spaces 
vector gauge fields either with spin-connections or with vielbeins, but it is also to show that 
expressing the gauge fields  with spin connections rather than with vielbeins makes the 
{\it spin-charge-family theory} transparent and correspondingly elegant, so that it is easier to
recognize that the origin of charges of the observed spinors, vector gauge fields, Higgs's
scalar and Yukawa couplings might really be in $(d-4)$ space, and that this explanation might 
show a possible next step beyond the {\it standard model}.

Let us remind the reader that the vector gauge fields, which carry the space index $m=(0,1,2,3)$, 
as well as the spinor fields, both observed in $d=(3+1)$, have in the Kaluza-Klein theories and in 
the {\it spin-charge-family} theory all the charges defined by the symmetry in $(d-4)$-%
dimensional space, while the (observed) dynamics of these fields is defined in $(3+1)$
space~\footnote{It is demonstrated on the special case in Ref.~\cite{hope} that the observed 
charges of spinors and of vector gauge fields (this is true also for the charges of the scalar fields)
originate in the lowest value of $M^{st}$, that is in $S^{st}$.}.

We present also the relation between the vielbeins and the spin connection fields for the scalar 
gauge fields - for the same symmetry of $d$-dimensional spaces % as for the vector gauge fields 
($g_{\sigma \tau} = \eta_{\sigma \tau}\, f^{-2}$ in $(d-4)$-dimensional space). While the vector 
gauge fields carry the space index $m=(0,1,2,3)$, the scalar gauge fields carry the space index
($s\ge 5$). Scalar gauge fields, carrying the space index $s=(7,8)$, manifest in $d=(3+1)$ as 
the  Higgs's scalar of the {\it standard model}, carrying the weak and the hyper charges 
$(\pm\frac{1}{2}, \mp\frac{1}{2}$),
respectively~\cite{norma2014MatterAntimatter,JMP2015,JMP2013,IARD2016}. Scalar gauge 
fields carry besides the properties defined by the space index (like there are the weak and hyper 
charges when the space index $s=(7,8)$) also the charges defined by the superposition of 
$S^{st}$ (superposition are determined by the symmetry of $(d-4)$ space).

There are spinor fields (and possibly also scalar gauge fields) which are responsible for curling 
($d-4$) space, forcing the space to manifest the required symmetry (Eqs.~(\ref{deltaxsigma})-
(\ref{Killing})). Consequently  vielbeins and spin connections of $(d-4)$ space reflect this 
symmetry and correspondingly these spinors (or possibly as well scalar gauge 
fields) do not enter into the relation of Eq. (\ref{omegaabe})~\footnote{If there are additional 
spinors, which do strongly influence the relation among vielbeins and spin connections, the spin 
connections are not any longer uniquely determined by the vielbeins, as demonstrated in 
Eq.~(\ref{omegaabe}). 
Then the symmetry of $(d-4)$ space might change further. It can happen, like in Ref.~\cite{hope,%
norma2004,hn2006}, that some of spinors stay massless after the break and the others do not, or like 
at the electroweak break when the symmetry of $(d-4)$ space breaks so that the weak and 
hyper charges break, keeping the electromagnetic charge unbroken~\cite{%
norma2014MatterAntimatter,IARD2016,n2012scalars,JMP2015,JMP2013}, while some scalars gain 
constant values  (called in the {\it standard model} the
nonzero vacuum expectation values) independent  of $(3+1)$ space coordinates.}.

Let us start with the action of the {\it spin-charge-family} theory~\cite{norma2014MatterAntimatter,%
n2012scalars,JMP2015,JMP2013,IARD2016}. In this simple action in an even dimensional space 
($d=2n$, $d>5$)~\footnote{In the {\it spin-charge-family} theory $d$ is chosen to be  $(13+1)$,
what makes the simple starting action in $d$ to manifest in $(3+1)$ in the low energy regime 
all the observed degrees of freedom, explaining all the assumptions of the {\it standard model} 
as well as other observed phenomena~\cite{norma2014MatterAntimatter,IARD2016,n2012scalars,%
JMP2015,JMP2013}.} 
fermions interact with the vielbeins $f^{\alpha}{}_{a}$ and the two kinds of the spin-connection 
fields - $\omega_{ab \alpha}$ and $\tilde{\omega}_{ab \alpha}$ -
the  gauge fields of $S^{ab} = \,\frac{i}{4} (\gamma^a\, \gamma^b
- \gamma^b\, \gamma^a)\,$ and $\tilde{S}^{ab} = \,\frac{i}{4} (\tilde{\gamma}^a\, 
\tilde{\gamma}^b - \tilde{\gamma}^b\, \tilde{\gamma}^a)$, respectively:
\begin{eqnarray}
{\cal A}\,  &=& \int \; d^dx \; E\;\frac{1}{2}\, (\bar{\psi} \, \gamma^a p_{0a} \psi) + h.c. +
%{\mathcal L}_{f} +  
\nonumber\\  
               & & \int \; d^dx \; E\; (\alpha \,R + \tilde{\alpha} \, \tilde{R})\,,%\nonumber\\
               %\end{eqnarray}
%
%\begin{eqnarray}
%{\mathcal L}_f &=& \frac{1}{2}\, (\bar{\psi} \, \gamma^a p_{0a} \psi) + h.c., 
%\nonumber\\
%p_{0a }        &=& f^{\alpha}{}_a p_{0\alpha} + \frac{1}{2E}\,
% \{ p_{\alpha}, E f^{\alpha}{}_a\}_-, 
%\nonumber\\  
%   p_{0\alpha} &=&  p_{\alpha}  - 
%                    \frac{1}{2}  S^{ab} \omega_{ab \alpha} - 
%                    \frac{1}{2}  \tilde{S}^{ab}   \tilde{\omega}_{ab \alpha},                   
%\nonumber\\ 
%R              &=&  \frac{1}{2} \, \{ f^{\alpha [ a} f^{\beta b ]} \;(\omega_{a b \alpha, \beta} 
%- \omega_{c a \alpha}\,\omega^{c}{}_{b \beta}) \} + h.c. \;, 
%\nonumber\\
%\tilde{R}      &=&  \frac{1}{2} \, \{ f^{\alpha [ a} f^{\beta b ]} \
%;(\tilde{\omega}_{a b \alpha,\beta} - 
%\tilde{\omega}_{c a \alpha} \,\tilde{\omega}^{c}{}_{b \beta})\} + h.c.\;, 
\label{wholeaction}
\end{eqnarray}
here $p_{0a } = f^{\alpha}{}_a\, p_{0\alpha} + \frac{1}{2E}\, \{ p_{\alpha},
E f^{\alpha}{}_a\}_- $, 
$ p_{0\alpha} =  p_{\alpha}  - \frac{1}{2} \, S^{ab}\, \omega_{ab \alpha} - 
                    \frac{1}{2} \,  \tilde{S}^{ab} \,  \tilde{\omega}_{ab \alpha} $,                    
$$R =  \frac{1}{2} \, \{ f^{\alpha [ a} f^{\beta b ]} \;(\omega_{a b \alpha, \beta} 
- \omega_{c a \alpha}\,\omega^{c}{}_{b \beta}) \} + h.c., $$  
$$\tilde{R}  =  \frac{1}{2} \, \{ f^{\alpha [ a} f^{\beta b ]} \;(\tilde{\omega}_{a b \alpha,\beta} - 
\tilde{\omega}_{c a \alpha} \,\tilde{\omega}^{c}{}_{b \beta})\} + h.c.~\footnote{Whenever 
two indexes are equal the summation over these two is meant.}. $$
 The action introduces two kinds of the Clifford algebra objects, $\gamma^a$ and
 $\tilde{\gamma}^a$,
\begin{eqnarray}
\label{twoclifford}
&&\{\gamma^a, \gamma^b\}_{+}= 2 \eta^{ab} = 
\{\tilde{\gamma}^a, \tilde{\gamma}^b\}_{+}\,.
\end{eqnarray}
$f^{\alpha}{}_{a}$ are vielbeins inverted to $e^{a}{}_{\alpha}$, Latin letters ($a,b,..$) denote
flat indices, Greek letters ($\alpha,\beta,..$) are Einstein indices,  $(m,n,..)$ and $(\mu,\nu,..)$ 
denote the corresponding indices in ($0,1,2,3$), $(s,t,..)$ and $(\sigma,\tau,..)$  denote the
corresponding indices in $d\ge5$:
\begin{eqnarray}
\label{vielfe}
e^{a}{}_{\alpha}f^{\beta}{}_{a} &=&\delta^{\beta}_{\alpha}\,, \quad
e^{a}{}_{\alpha}f^{\alpha}{}_{b}= \delta^{a}_{b}\,,
\end{eqnarray}
$E =det(e^{a}{}_{\alpha})$~\footnote{This definition of the vielbein and the inverted vielbein is 
general, no specification about the curled space is assumed yet. In Eq.(\ref{ef}) vielbeins are 
specified for the case that $(d-4)$ space is curled, Eqs.~(\ref{fespecial},\ref{Estspecial}), while 
$f^{\sigma}{}_{m}$ determines vector gauge fields $\Omega^{st}{}_{m}$ as presented in 
Eq.~(\ref{feOmega}).}.  
The action ${\cal A} $ offers the explanation for the origin and all the properties of the observed 
fermions (of the family members and families), of the observed vector gauge fields, of the Higgs's
scalar and of the Yukawa couplings, explaining the origin of the matter-antimatter asymmetry,  the 
appearance of the dark matter and predicts new scalars and new family %and a new gauge field 
to be observed at the LHC~(\cite{norma2014MatterAntimatter,IARD2016} and the 
references therein). 

The spin connection fields and the vielbeins are related fields and,  if there are no spinor (fermion) 
sources present (both kinds of, the one of $S^{ab}$ and the one of $\tilde{S}^{ab}$) the spin 
connection fields are expressible with the vielbeins. 
In Ref.~\cite{n2012scalars} (Eq.~(C9)) the expressions relating the spin connection fields of 
both kinds with the vielbeins and the spinor sources are presented.

 We present below the relation among the $\omega_{ab\alpha} $ fields and the vielbeins 
% with no sources present, which is relevant for our 
%discussions 
(\cite{norma2004}, Eq.~(6.5), where the relation 
$$e^{a}{},_{\alpha} +
 \omega^{a}{}_{b \alpha} \, e^{b}{}_{\beta} - 
\Gamma^{\alpha'}{}_{\beta \alpha} \, e^{a}{}_{\alpha'}=0 $$ 
is used, with $ \Gamma^{\alpha}{}_{\beta \gamma} =\frac{1}{2} \, g^{\alpha \delta}$ 
$ (g_{\beta \delta},_{\gamma}+ g_{\gamma \delta},_{\beta} - g_{\beta \gamma},_{\delta})$ ), 
(\cite{JMP2015}, Eq.~(C9)).
\begin{eqnarray}
\label{omegaabe}
\omega_{ab}{}^{e} &=& 
 \frac{1}{2E} \{   e^{e}{}_{\alpha}\,\partial_\beta(Ef^{\alpha}{}_{[a} f^\beta{}_{b]} )
      - e_{a\alpha}\,\partial_\beta(Ef^{\alpha}{}_{[b}f^{\beta e]})
  \nonumber\\
                  & &  \qquad\qquad  {} - e_{b\alpha} \partial_\beta (Ef^{\alpha [e} f^\beta{}_{a]})\}
                     \nonumber\\
                  &+& \frac{1}{4}   \{\bar{\Psi} (\gamma^e \,S_{ab} - 
 \gamma_{[a}  S_{b]}{}^{e} )\Psi \}  \nonumber\\
                  &-& \frac{1}{d-2}  
   \{ \delta^e_{a} [
\frac{1}{E}\,e^d{}_{\alpha} \partial_{\beta}
             (Ef^{\alpha}{}_{[d}f^{\beta}{}_{b]})
                        + \bar{\Psi} \gamma_d  S^{d}{}_{b} \,\Psi ] \nonumber\\  & & \qquad {}
     - \delta^{e}_{b} [
            \frac{1}{E} e^{d}{}_{\alpha} \partial_{\beta}
             (Ef^{\alpha}{}_{[d}f^{\beta}{}_{a]} )
            + \bar{\Psi} \gamma_{d}  S^{d}{}_{a}\, \Psi ]
\}\,. \nonumber\\ \label{omegas} 
                        \end{eqnarray}
When the gauge vector and  scalar fields in $d=(3+1)$ are studied, with the charges originating in 
$(d-4)$-dimensional space, 
%with no presence of the gravitational field in $d=(3+1)$,  
the denominator $\frac{1}{d-2}$ must be replaced by  $\frac{1}{(d-4)-2}$.
%!!The expression for the spin connection fields carrying family quantum numbers - 
%$\tilde{\omega}_{ab}{}^e$ - is in the case that 
%there are no spinor sources identical with the right hand side of Eq.~\ref{omegaabe}.)
One notices that if there are no spinor sources present, carrying the spinor quantum numbers 
$S^{ab}$, then $\omega_{abc}$  is completely determined by the vielbeins (and so is 
$\tilde{\omega}_{abc}$ if there are no spinor sources present carrying $\tilde{S}^{ab}$).
 Eq.~(\ref{omegaabe}) manifests that the last terms with $\delta^{e}_{a}$ and 
$\delta^{e}_{b}$ do not contribute when the vector gauge fields  $\omega_{st}{}^{m}$, 
$(s,t) = (5,6,\dots, d)$ and $m=(0,1,2,3)$,  are under consideration.

We demonstrate in this paper, Sect.~\ref{proof}, that in the spaces with the maximal number 
of the Killing vectors~(\cite{mil}, p. (331--340)) and with no spinor sources present (which would
change the symmetry of $(d-4)$ space),  the vielbeins 
$f^{\sigma}{}_{m}$ and the spin connections $\omega_{stm}$ are in the Kaluza-Klein 
theories~\cite{zelenaknjiga,mil} related. We find, Eqs.~(\ref{feOmega},\ref{omegaabe1m1}): 
$ f^{\sigma}{}_{m}= $
$- \frac{1}{2}\, E^{\sigma}_{st}  \,\omega^{st}{}_{m} (x^{\nu})\,$. When the vector 
gauge fields are 
superposition of the spin connection fields  ($A^{Ai}_{m} =\sum_{s,t} c^{Ai}{}_{st}\, 
\omega^{st}{}_{m}$), the relations among the vielbeins and spin connections are correspondingly: 
$f^{\sigma}{}_{m}= \sum_{A}\,\vec{\tau}^{A\sigma}\, \vec{A}^{A}_{m}$, 
as presented in Eqs.~(\ref{fmagen}-\ref{fmagennew}).
Spinors, vector gauge fields and scalar gauge fields, manifested in $d=(3+1)$ as dynamical fields, 
can be treated as weak fields, which do not influence the symmetry of $(d-4)$ space. 
When these fields start to be strong the symmetry of the curled space changes. (Let us
mention that, for example, scalar fields at the electroweak break do break the symmetry of
$(d-4)$ space.)

Since the vielbeins $f^{\alpha}{}_{a} $ and inverted vielbeins $e^{a}{}_{\alpha}$ 
(Eq.~(\ref{vielfe})) appear in the metric tensor as a product ($g^{\alpha \beta}=
f^{\alpha}{}_{a} f^{\beta a}, \, g_{\alpha \beta}= e^{a}{}_{\alpha} e_{a \beta} $),
also tensors of the vector gauge fields appear in $d=(3+1)$ in the curvature $R^{(d)}$ 
as it is expected for the vector gauge fields,
Eqs.~(\ref{fmagennew},\ref{Lagrange},\ref{actionvg}): 
$$R^{(d)} =R^{(4)} + 
R^{(d-4)} - \frac{1}{4} g_{\sigma \tau} E^{\sigma}{}_{st}
 E^{\tau}{}_{s' t'} \, F^{s t}{}_{\mu \nu} \, F^{s' t' \mu \nu}.$$ 

We demonstrate in Sect.~\ref{scalars} that also spin connection fields $\omega^{st}{}_{s'}$ 
(with the index $s'$ from $(d-4)$ space, and accordingly scalar with respect to $(3+1)$ space)  
are uniquely expressible by vielbeins, Eqs.~(\ref{Rscalarspecial1},\ref{Rscalarspecial2}), as
long as the curled space has large enough symmetry. Consequently also the superposition of 
the scalar spin connection fields are expressible with the vielbeins.

\section{Proof that spin connections and vielbeins lead to the same vector gauge fields in $(3+1)$-%
dimensional space-time}
\label{proof}

We discuss relations between  spin connections and vielbeins when
% there are no spinor sources present and 
space in $(d - 4)$ demonstrates the desired isometry 
%(keeping the form of the metric in $(d-4)$ space unchanged)
 in order to prove that both ways, either using vielbeins or spin connections, lead to equivalent 
vector gauge fields in $(3+1)$.

We point out that spin connections manifest (charges and properties of) vector gauge 
fields more transparently (and elegantly) than vielbeins~\footnote{  
In addition: At low energies there are superposition of spins of spinors, which manifest charges 
of spinors in $(3+1)$, and there are superposition of $S^{st}$ acting on superposition of spin 
connection fields which manifest as the charges of vector (and scalar) gauge fields (vectors 
manifest in addition to charges - originating $(d-4)$ - in $SO(3+1)$ the $SU(2) \times SU(2)$ 
spin structure, while scalars carry besides  charges - originating $(d-4)$,  of the same 
origin as there are charges of vector gauge fields -  also the properties 
defined by the space index in $(d-4)$.
All these support the idea that the origin of vector (as well as scalar) gauge fields might indeed 
be in higher dimensional space.}. 

% 

%One of the authors (N.S.M.B.) spent a lot of time and effort (Ref. 11.), also the second 
% author collaborated (Refs 11), to prove that the presence of the spinor fields can 
%take care of the spin connection fields which together with the vielbeins makes spinors of one 
%handedness massless when the space (d-4) breaks.. Or in short: Both – spin connections and 
%vielbeins are needed that the Kaluza-Klein theories and correspondingly also the spin-charge-family 
%theory manifest in the low energy regime the observed symmetries.

Let ($d-4$) space manifest the rotational symmetry, determined by the infinitesimal coordinate 
transformations of the kind
\begin{eqnarray}
\label{deltaxsigma}
x'^{\mu} &=& x^{\mu}\,, \nonumber\\
 x'^{\sigma} &=& x^{\sigma} +\varepsilon^{st}(x^{\mu})\, E^{\sigma}_{st} (x^{\tau})=
 x^{\sigma} - i \varepsilon^{st} (x^{\mu})\, M_{st} \,x^{\sigma}\,,
\end{eqnarray}
where $M^{st}= S^{st} + L^{st}$, $L^{st}=x^s p^t-x^t p^s$, $S^{st}$ concern internal 
degrees of freedom of boson and fermion fields,  $\{M^{st}, 
M^{s't'}\}_{-} = i (\eta^{s t'} M^{ts'} + \eta^{ts'} M^{st'} - \eta^{s s'} M^{tt'} - 
\eta^{t t'} M^{ss'})$~\footnote{While $L^{st}$ act on coordinates,  $S^{st}$ act on  
spinor fields, on vector gauge fields (they are superposition of $\omega_{stm}$, 
$(s,t)$ belonging to $(d-4)$ space, $m$ to $(3+1)$ space) and on scalar gauge fields (they are
 superposition of 
$\omega_{s t t'}$, $(s,t, t')$ belonging to $(d-4)$ space), the charges of which originate in higher 
dimensional space and correspondingly $S^{st}$ act on their charges (which are the 
superposition of $S^{st}$). For example, ${\cal S}^{ab}$ act on gauge fields~\cite{IARD2016}
as follows: ${\cal S}^{ab} \, A^{d\dots e \dots g} = i \,(\eta^{be}\,A^{d\dots a \dots g} - 
\eta^{ae} \,
A^{d\dots b \dots g}$).}. 
From Eq.~(\ref{deltaxsigma}) it follows 
\begin{eqnarray}
\label{deltaxsigma1}
-i \, M_{st}\,  x^{\sigma} &=&  E^{\sigma}_{st} = x_{s}\, f^{\sigma}{}_{t} - 
x_{t}\, f^{\sigma}{}_{s}\,,\nonumber\\
E^{\sigma}_{st} &=& (e_{s  \tau}\, f^{\sigma}{}_{t} - e_{t  \tau}\, f^{\sigma}{}_{s}) x^{\tau}
\,,\nonumber\\
M_{st}{}^{\sigma}: &=& i E^{\sigma}_{st}\,,
\end{eqnarray}
and correspondingly: $M_{st} =  E^{\sigma}_{st}\,  p_{\sigma}$. 
One derives, when taking into account  Eq.~(\ref{deltaxsigma1}) and the commutation relations 
among generators of the infinitesimal rotations, the equation for the Killing vectors 
$E^{\sigma}_{s t}$
%
%\begin{eqnarray}
\begin{multline}
\label{comE}
E^{\sigma}_{s t} p_{\sigma}  E^{\tau}_{s' t'} p_{\tau} - 
E^{\sigma}_{s 't'} p_{\sigma}  E^{\tau}_{s t} p_{\tau} = \\
- i (\eta_{s t'} E^{\tau}_{t s'}  + \eta_{t s'} E^{\tau}_{s t'} - \eta_{s s'} E^{\tau}_{t t'} 
- \eta_{t t'} E^{\tau}_{s s'}) p_{\tau}\,,  
\end{multline}
%\end{eqnarray}
%
and the Killing equation
\begin{eqnarray}
\label{Killing}
&&D_{\sigma} E_{\tau s t}  + D_{\tau} E_{\sigma s t} =0\,,\nonumber\\
&&D_{\sigma} E_{\tau s t} = \partial_{\sigma}  E_{\tau s t}  - \Gamma^{\tau'}{}_{\tau \sigma}
 E_{\tau' s t}\,. 
\end{eqnarray}
Let the corresponding background field ($g_{\alpha \beta} = e^{a}{}_{\alpha} \,e_{a\beta}$) 
be
\begin{eqnarray}
\label{ef}
e^{a}{}_{\alpha}=
\begin{pmatrix} \delta^{m}{}_{\mu} & e^{m}{}_{\sigma}=0 \\
 e^{s}{}_{\mu} & e^s{}_{\sigma} 
\end{pmatrix}
\,, \quad 
f^{\alpha}{}_{a} =
\begin{pmatrix} \delta^{\mu}{}_{m}  & f^{\sigma}{}_{m} \\
0= f^{\mu}{}_{s} & f^{\sigma}{}_{s}\,,  
\end{pmatrix}
\,,
\label{fe}
\end{eqnarray}
so that  the background field in $ d=(3+1)$ is flat.
% We asume that the vielbein components $e^{s}{}_{\mu}$ and $f^{\sigma}{}_{m}$ depend 
%only on $x^{\mu}$ and $x^{\sigma}$ as presented in Eq.~(\ref{gctf}). Check all the 
%symmetries carefully.
 From $e^{a}{}_{\mu}f^{\sigma}{}_{a}=$
$\delta^{\sigma}_{\mu}=0 $ it follows
\begin{eqnarray}
\label{fe1}
e^{s}{}_{\mu} &=& -\delta^{m}_{\mu} \, e^{s}{}_{\sigma}\, f^{\sigma}{}_{m} \,.  
\end{eqnarray}
This leads to
\begin{equation}
g_{\alpha \beta} =
\begin{pmatrix}
 \eta_{m n} +  f^{\sigma}{}_{m}  f^{\tau}{}_{n}  
e^{s}{}_{\sigma} e_{s \tau}\;\;\; & -f^{\tau}{}_{m} e^{s}{}_{\tau} e_{s\sigma}\\
-f^{\tau}{}_{n} e^s{}_{\tau} e_{s \sigma} & e^s{}_\sigma e_{s\tau}
\end{pmatrix}\,, 
\label{gmdown}
\end{equation}
and
\begin{equation}
g^{\alpha \beta} =
\begin{pmatrix}
 \eta^{m n}\quad \quad\quad & f^{\sigma}{}_{m}\\
f^{\sigma}{}_{m} \quad \quad\quad & f^{\sigma}{}_{s}  f^{\tau s} +
 f^{\sigma}{}_{m}  f^{\tau m}
\end{pmatrix}\,.
\label{gmup}
\end{equation}
We have: $ \Gamma^{\tau'}{}_{\tau \sigma} =\frac{1}{2} \, g^{\tau' \sigma'}$ 
$ (g_{\sigma \sigma'},_{\tau}+ g_{\tau \sigma'},_{\sigma} - g_{\sigma \tau},_{\sigma'})$.

One can check properties of $f^{\sigma}{}_{m} \,\delta^{m}_{\mu}$ under general coordinate
transformations: $x'^{\mu} =x'^{\mu}(x^{\nu})$, $x'^{\sigma}=$ $ x'^{\sigma}(x^{\tau})$, 
$\left(g'_{\alpha \beta} = \frac{\partial x^{\rho}}{\partial x'^{\alpha}} 
 \frac{\partial x^{\delta}}{\partial x'^{\beta}} g_{\rho \delta} \right)$, 
\begin{eqnarray}
\label{gctf}
f'^{\sigma}{}_{m}\delta^{m}_{\mu} &=&\frac{\partial x^{\nu}}{\partial x'^{\mu}}
\frac{\partial x'^{\sigma}}{\partial x^{\tau}} f^{\tau}{}_{\nu}.
%-  \frac{\partial x'^{\sigma}}{\partial x^{\nu}}) \,.  
\end{eqnarray}
Let us introduce the vector gauge field $\Omega^{st}{}_{m} (x^{\nu})$, which depends only 
on the coordinates in $d=(3+1)$, as  follows
\begin{eqnarray}
 f^{\sigma}{}_{m}: &=& - \frac{1}{2}\, E^{\sigma}_{st}  
\,\Omega^{st}{}_{m} (x^{\nu})\,,
\label{feOmega}
\end{eqnarray}
with $ E^{\sigma}_{st}= -i M_{st}{}^{\sigma} $ defined in Eq.~(\ref{deltaxsigma1}).
%%%%%%%%%%%%%%%%
$f^{\sigma}{}_m$ depends on the (3+1) coordinates through  $\Omega^{st}{}_{m}$ and
on $(d-4)$ coordinates through $ E^{\sigma}_{st}$~\footnote{These gauge fields 
$\Omega^{st}{}_{m}$ are in the low energy regime weak in comparison with the fields which 
force the space to curl. The influence of these gauge fields and correspondingly of 
$f^{\sigma}{}_m$ on the equations of motion of vielbeins and spin connections in the higher 
dimensional space can be assumed as negligible, as is the case for weak spinor sources (this is
the usual  procedure in problems in classical or quantum mechanics when weak perturbation is 
put into strong fields, this prosedure is assumed also in the Kaluza-Klein theories).}. 
%%%%%%%%%%%%%%%%
%CHECK below again, also factor2 and -
From Eqs.~(\ref{gctf},\ref{feOmega})  the transformation properties of 
$\Omega^{st}{}_{m} $ under the coordinate transformations of Eq.~(\ref{deltaxsigma}) follow. 
%
%\begin{eqnarray}
%-  E^{\sigma st} \, \delta_0 \,\Omega_{st m} &=& -  E^{\sigma st} {\bf \{}  -
% \varepsilon_{s t}{}_{,m}  +i 2 (\varepsilon_{s}{}^{s'}\,\Omega_{s't m} -  
%\varepsilon_{t}{}^{s'}\,\Omega_{s's m})  {\bf \}}\,.
%\label{feOmega1}
%\end{eqnarray}
%
%

If we look for the transformation properties of the superposition of the fields $\Omega_{st m}$, 
let say 
$$ {\cal A}^{Ai}{}_{m}=\sum_{s,t}\, c^{Ai st} \Omega_{st m},$$ which are the gauge 
fields of $\tau^{Ai}$ (with the commutation relations $\{\tau^{Ai}, \tau^{Bj}\}_{-} = 
i \delta^{A}_{B} \,f^{Aijk} \tau^{Ak}$, where $\tau^{Ai}=\sum_{s,t} c^{Ai}{}_{st} M^{st}$ 
and $f^{Aijk}$ are the structure constants of the corresponding gauge groups), 
under the  coordinate transformations of Eq.~(\ref{deltaxsigma}), one finds 
%
%\begin{eqnarray}
$ \delta_0 \,{\cal  A}^{Ai}{}_{m} $ $=  \varepsilon^{Ai}{}_{,m} + i f^{A ijk}\, {\cal A}^{Aj}_{m}\,
\varepsilon^{Ak}$. %\nonumber
%\label{feA}
%\end{eqnarray}
%

Let us make  a choice of $f^{\sigma}{}_{s}$
\begin{eqnarray}
f^{\sigma}{}_{s}&=& f\, \delta^{\sigma}_{s}\,,  \nonumber \\
e^{s}{}_{\sigma} &=& f^{-1} \delta^{s}{}_{\sigma}\,, 
\label{fespecial}
\end{eqnarray}
for which $E^{\sigma}_{st}$ is equal to
\begin{eqnarray}
E^{\sigma}_{st}&=& (\eta_{s \tau} \delta^{\sigma}_{t} - \eta_{t \tau} \delta^{\sigma}_{s})
x^{\tau}\,,
\label{Estspecial}
\end{eqnarray}
%
%and $E^{\sigma}_{Ai}$ is equal to
%
%\begin{eqnarray}
%E^{\sigma}_{Ai}&=& \sum_{s,t} c^{st}{}_{Ai}\,(\eta_{s \tau} \delta^{\sigma}_{t} - 
%\eta_{t \tau} \delta^{\sigma}_{s})
%x^{\tau}\,,
%\label{EAispecial}
%\end{eqnarray}
%
%
solving the Killing equation (\ref{Killing}) if $f$ is the scalar function of the coordinates. 
Let us put the expression for $f^{\sigma}{}_{m}$,  Eq.~(\ref{feOmega}), into 
Eq.~(\ref{omegaabe}) to see the relation among $\omega_{stm}$ and $f^{\sigma}{}_{m}$. 
One finds 
\begin{eqnarray}
\label{omegaabe1m}
\omega_{stm} &=&  \frac{1}{2E} \{   f^{\sigma}{}_{m}\,[e_{t \sigma}\partial_\tau
(Ef^{\tau}{}_{s})  - e_{s\sigma}\,\partial_\tau (Ef^{\tau}{}_{t})]
  \nonumber\\
                  & & + e_{s\sigma} \partial_\tau [E (f^{\sigma}{}_{m}
 f^{\tau}{}_{t} - f^{\tau}{}_{m}  f^{\sigma}{}_{t})]\nonumber\\
 & & - e_{t\sigma} \partial_\tau [E
 (f^{\sigma}{}_{m}  f^{\tau}{}_{s} - f^{\tau}{}_{m}  f^{\sigma}{}_{s})] \}\,.
  \end{eqnarray}
(Since we study only the relation between vielbeins and spin connections when there are no 
spinor sources present, either weak or strong, the term $\psi^{\dagger} \gamma^0 
\gamma^m S_{st} \psi$ is dropped. Studying problems with the weak spinor sources present
would only slightly complicate the problem, while it would make the proof less transparent.)
 
Using the inverse vielbeins 
$e^{s}{}_{\sigma} = $ $f^{-1} \delta^{s}{}_{\sigma}$ and  
$$det(e^{s}{}_{\sigma}) =E =
f^{-(d-4)} $$ (Eq.~(\ref{fe})) and taking $\Omega_{stm} =\Omega_{stm} (x^n) $, as assumed 
above, it follows (after using Eq.~(\ref{feOmega}) and recognizing that 
$f^{\sigma}{}_{m} = - \frac{1}{2}\, (e_{s' \tau'} f^{\sigma}{}_{t'} - 
e_{t' \tau'} f^{\sigma}{}_{s'})\, x^{\tau'} \, \Omega^{s't'}{}_{m}$ )
\begin{eqnarray}
\label{omegaabe1m1}
\omega_{stm} &=& \frac{1}{2}\,(\eta_{s \sigma}\,\delta^{\tau}_{t} - 
\eta_{t \sigma}\,\delta^{\tau}_{s})\, \partial_{\tau} f^{\sigma}_{ m}\,,
\nonumber\\
\omega_{stm} &=&\Omega_{stm}\,.
\end{eqnarray}
It is therefore proven for the vielbeins 
$$ f^{\sigma}{}_{m}= - \frac{1}{2}\, E^{\sigma}_{st}  
\,\omega^{st}{}_{m} (x^{\nu}),$$ Eq.~(\ref{feOmega}), where in $d\ge5$ vielbeins solve the 
Killing equation~(\ref{Killing}), that the spin connections determine the gauge vector fields in 
$d=(3+1)$.

{\it Statement:} {\it Let the space with $s\ge 5$ have the symmetry allowing the infinitesimal 
transformations of the kind}
\begin{eqnarray}
\label{deltaxsigmagen}
x'^{\mu} &=& x^{\mu}\,, \quad 
 x'^{\sigma} = %x^{\sigma} \varepsilon^{st}(x^{\mu})\, E^{\sigma}_{st} (x^{\tau})=
 x^{\sigma} - i  \, \sum_{A,i, s,t} \varepsilon^{Ai} (x^{\mu})\, c_{Ai}{}^{st }M_{st} \,
x^{\sigma}\,,\nonumber\\
\end{eqnarray}
{\it then the vielbeins} $f^{\sigma}{}_{m}$ in Eq.~(\ref{fe}) {\it manifest in} $d=(3+1)$ {\it the 
vector gauge fields} ${\cal A}^{Ai}_{m}$
\begin{eqnarray}
\label{fmagen} 
f^{\sigma}{}_{m}&=& \sum_{A}\,\vec{\tau}^{A\sigma}\, \vec{\cal{A}}^{A}_{m} \,, 
\end{eqnarray}
{\it where} 
% Check $-i$ at 656 or there must be $i$on line 658 and 659
\begin{eqnarray}
\label{taua} 
\tau^{Ai} &=& \sum_{s,t}\, c^{Ai}{}_{st}\,M^{st} \,,\nonumber\\
\{\tau^{Ai}, \tau^{Bj}\}_{-}& = & i f^{Aijk} \tau^{Ak}\, \delta^{A B}\,,\nonumber\\
\vec{\tau}^{A} &=& \vec{\tau}^{A \sigma}\, p_{\sigma} =
 \vec{\tau}^{A\sigma}{}_{\tau}\,  x^{\tau}\,p_{\sigma}\,\nonumber\\
 \tau^{Ai \sigma} &=& \sum_{s,t}\, -i 
c^{Ai}{}_{st}\,M^{st\sigma}\nonumber\\
 &=& \sum_{s,t}\, c^{Ai}{}_{st}\,  (e_{s  \tau}\, f^{\sigma}{}_{t} - e_{t  \tau}\,
 f^{\sigma}{}_{s}) x^{\tau}= E^{\sigma}_{Ai}\,,\nonumber\\
{\cal A}^{Ai}_{m}&=& \sum_{s,t} \,c^{Ai}{}_{st} \, \omega^{st}{}_{m}\,.
\end{eqnarray}
The relation between $ \omega^{st}{}_{m}$ and 
vielbeins is determined by Eq.~(\ref{omegaabe1m}).

We have to express  $ A^{Ai}{}_{m}=\sum_{s,t}\, c^{Ai st} \,\omega_{st m}$ using  
Eq.~(\ref{omegaabe1m}).
Then it is not difficult to see 
%that, if using  Eq.~(\ref{omegaabe1m}) to find the relation between 
%$$ A^{Ai}_{m} = \sum_{s,t} \,c^{Ai}{}_{st} \, \omega^{st}{}_{m}$$ and $f^{\sigma}{}_{m}$
%of Eq.~(\ref{fmagen}), 
that we end up with the relation 
\begin{eqnarray}
\label{AcalA} 
A^{Ai}_{m}&=&{\cal A}^{Ai}_{m} \,, 
\end{eqnarray}
leading to the equation 
\begin{eqnarray}
\label{fmagennew} 
f^{\sigma}{}_{m}&=& \sum_{A}\,\vec{\tau}^{A\sigma}\, \vec{A}_{m} \,. 
\end{eqnarray}

The Lagrange function for these vector gauge fields follows from the curvature in $d$ dimensional
space 
$$R= R^{\alpha}{}_{\beta \alpha \gamma} g^{\beta \gamma},$$ after using 
 Eqs.~(\ref{gmdown},\ref{gmup}) in the relation for  $ \Gamma^{\alpha}{}_{\beta \gamma} 
=\frac{1}{2}  
g^{\alpha \delta}\, (g_{\gamma \delta},_{\beta} + g_{\beta \delta},_{\gamma} - $
$g_{\beta \gamma},_{\delta})$ and after taking into account this relation in the Riemann tensor
$R^{\alpha}{}_{\beta \gamma \delta} =\Gamma^{\alpha}{}_{\beta [\gamma},_{\delta]}$
$+\Gamma^{\alpha}{}_{\alpha' [\gamma}  \Gamma^{\alpha'}{}_{\beta \delta]}$, where 
${},_{\delta}$  denotes the derivative with respect to $x^{\delta}$
 ($\frac{\partial}{\partial x^{\delta}}$) and the
parentheses require antisymmetrization of the two indexes.

For a flat  four dimensional space ($R^{(4)}=0$) it follows for the curvature~(\cite{mil}, Eq.~(10.41))
\begin{eqnarray}
\label{Lagrange}
R^{(d)} &=&% R^{(4)} + 
R^{(d-4)} - \frac{1}{4} g_{\sigma \tau} E^{\sigma}{}_{st}
 E^{\tau}{}_{s' t'} \, F^{s t}{}_{m n} \, F^{s' t' m n}\,,\nonumber\\
 F^{s t}{}_{m n }&=& \partial_{m} A^{st}_{n} - \partial_{n} A^{st}_{m}
-f^{st}{}_{s' t' s" t"} \, A^{s' t'}_{m} \,A^{s'' t"}_{n}\,,\nonumber\\
f^{\sigma}{}_{m} &=& -\frac{1}{2} E^{\sigma}{}_{s t} \, \omega^{s t}{}_{\mu}\,
f^{\mu}{}_{m}\,,\nonumber\\
E^{\sigma}{}_{st} &=&-i \,M^{st} x^{\sigma} = (e_{s \tau} f^{\sigma}{}_{ t}- 
e_{t \tau} f^{\sigma}{}_{s}) x^{\tau}\,,
\end{eqnarray}
where $ R^{(d-4)} $ determines the curvature in $(d-4)$ dimensional space and 
$f^{st}{}_{s' t' s" t"} $ can be obtained from the commutation relations
$\{M^{st}, 
M^{s't'}\}_{-} = i (\eta^{s t'} M^{ts'} + \eta^{ts'} M^{st'} - \eta^{s s'} M^{tt'} - 
\eta^{t t'} M^{ss'})$.
Vielbein $f^{\sigma}{}_{m}$ simplifies, when $f^{\sigma}{}_{s} = f \delta^{\sigma}_{s}$ and 
$d=(3+1)$ is a flat space, to $f^{\sigma}{}_{m} = \omega^{\sigma}{}_{\tau m}\,x^{\tau}$.

When $(d-4)$ space manifests the symmetry of Eq.~(\ref{fmagen})
($f^{\sigma}{}_{m}= \sum_{A}\,\vec{\tau}^{A\sigma}\, \vec{A}^{A}_{m})$  and $d=(3+1)$
is a flat space, the curvature  $R^{(d)}$ becomes equal to~\cite{mil} (Eq. (10.41))~\footnote{
Ref.~\cite{MatejPavsic}, Sect.~5.3,  deriving the Lagrange function for the gauge fields by using the 
Clifford algebra space, allows both, the curvature $R$ and its quadratic form $R^2$, Eq.~(240). } 
\begin{eqnarray}
\label{actionvg}
R^{(d)} &=& R^{(d-4)} \nonumber\\
 & & - \frac{1}{4}\, \sum_{\substack{A, i, A', i', \\
  \sigma, \tau,\mu,\nu}}
 g_{\sigma \tau} E^{\sigma}{}_{A i}
 E^{\tau}{}_{A' i'} \, F^{Ai}{}_{mn} F^{A' i' \,mn}\,,\nonumber\\
F^{Ai}{}_{mn} &=& \partial_{m} A ^{Ai}_{n}- \partial_{n} A ^{Ai}_{m}
- i f^{Aijk} \, A^{Aj}_{m} \,A^{Ak}_{n}\,,\nonumber\\
A^{Ai}_{m}&=& \sum_{s, t} \,c^{Ai}{}_{s t}\, \omega^{s t}{}_{m}\,,\nonumber\\
\tau^{Ai} &=&  \sum_{s, t}\, c^{Ai s t} \,M_{s t}\,,
\end{eqnarray}
with $ E^{\tau}{}_{A i }$ defined in Eq.~(\ref{taua}).
 
The integration of the action $ \int \,E \,d^{4} x \, d^{(d-4)} x \,R^{(d)}$ over an even
dimensional $(d-4)$ space 
%(in which only even functions of the coordinates $x^{\sigma}$ give nonzero contributions)
leads to the well known effective action for the vector gauge fields in $d=(3+1)$ space: 
$\int E' \,d^{4} x \, \{ - \frac{1}{4} \,\sum_{A,i,m,n} F^{A i}{}_{m n} \, F_{A i}{}^{mn} $, 
where $E'$ is determined by the gravitational field in $(~3~+~1~)$ space ($E'=1$, if $(3+1)$ 
space is flat). 
%
%\begin{eqnarray}
%\label{Lagrangesup}
%R^{(d)} &=&% R^{(4)} + 
%R^{(d-4)} - \frac{1}{4} g_{\sigma \tau} E^{\sigma}{}_{A i}
% E^{\tau}{}_{A i} \, F^{A i}{}_{\mu \nu} \, F^{A i \mu \nu}\,,\nonumber\\
% F^{A i}{}_{\mu \nu}&=& \partial_{\mu} A^{A i}_{\nu} - \partial_{\nu} A^{A i}_{\mu}
%-f^{A i j k}  A^{A j}_{\mu} \,A^{A k}_{\nu}\,.
%%\nonumber\\
%f^{\sigma}{}_{m} &=& -\frac{1}{2} E^{\sigma}{}_{(s t)} \, \omega^{s t}{}_{\mu}\,
%f^{\mu}{}_{m}\,,\nonumber\\
%E^{\sigma}{}_{(st)} &=& M^{st} x^{\sigma} = (e_{s \tau} f^{\sigma}{}_{ t}- 
%e_{t \tau} f^{\sigma}{}_{s}) x^{\tau}\,.
%\end{eqnarray}
%
All the vector gauge fields (manifesting in $d=(3+1)$, $x^{m}$ are coordinates in a flat $(3+1)$ 
space) are superposition of the spin connection fields: $A^{Ai}_{m} = \sum_{s,t}\, 
c^{Ai}{}_{st}\, \omega^{st}{}_{m}$, the charges of which ($\tau^{Ai}=\sum_{s,t} 
c^{Ai}{}_{st}\,S^{st}$) are determined by the symmetry of $(d-4)$ space. 

This completes the proof of the above statement, that the vielbeins $f^{\sigma}{}_{m}$, 
$\sigma=(5,6,\dots,d)$, $m=(0,1,2,3)$, are expressible with the spin connection fields 
$\omega_{s t m}$: $f^{\sigma}{}_{m}= \sum_{A,i,s,t,}\,\tau^{A i \sigma}\, c_{A i}{}^{s t} 
\, \omega_{s t m}$~\footnote{In general not only $S^{st}$ but the total angular momentum 
$M^{st}$ ($=$  $L^{st} + S^{st}$) contribute to the charges of the vector gauge fields,
manifesting in this case higher charges~\cite{hope,norma2004,hn2006}, but this is not what manifests in 
the low energy region.}.

Since vector gauge fields are direct ($c_{A i}{}^{s t} $ are complex numbers) superposition of 
spin connection fields, the spin connection fields offer an elegant and transparent description of the 
vector gauge fields. This is what the {\it spin-charge-family} theory is using.

In Subsect.~\ref{demonstration} we demonstrate the connection among the spin connection 
fields $\omega_{s t m}$ and the vielbeins  $f^{\sigma}{}_{m}$ when  ($d-4$) space manifests
the $SU(2)\times SU(2)$ symmetry. Generalization to any symmetry in  $(d-4)$ space goes 
in a similar way, leading to the corresponding expressions for the vector gauge fields in $d=(3+1)$.

\subsection{Vector gauge fields $SU(2) \times SU(2)$ as the superposition of the spin connections}
\label{demonstration}

Let us demonstrate the statement that all the vector gauge fields are superposition of the spin 
connection fields in the case, when the space of the symmetry $SO(7,1)$  breaks into 
$SO(3,1)\times SU(2)\times SU(2)$.

One finds the coefficients $c^{Ai}{}_{st}$ for the two $SU(2)$ generators,  $\tau^{1i}$
 $= \sum_{s,t} c^{1i}{}_{st}\, M^{st}$ and $\tau^{2i}$ $=\sum_{s,t} c^{2i}{}_{st}\,
 M^{st}$ by requiring the 
commutation relations $\{\tau^{Ai}, \tau^{Bj}\}_{-}=\delta^{A B} f^{A ijk}\, 
\tau^{Ak}$,
\begin{eqnarray}
\label{tau1}
\vec{\tau}^1 &=& \frac{1}{2} \,(M^{58} - M^{67}, M^{57} + M^{68}, M^{56} - M^{78})
\nonumber\\
\vec{\tau}^{2} &=& \frac{1}{2}\, (M^{58} + M^{67}, M^{57} - M^{68},M^{56} + M^{78})\,,
\end{eqnarray}
while one finds coefficients $c^{1i}{}_{st}$ and $c^{2i}{}_{st}$ for the corresponding gauge 
fields 
\begin{eqnarray}
\label{A1A2}
\vec{A}^{1}_{a} &=& \frac{1}{2} \,(\omega_{58a} - \omega_{67a},\, \omega_{57a} +
\omega_{68a},\, \omega_{56a} - \omega_{78a})
\nonumber\\
\vec{A}^{2}_{a} &=& \frac{1}{2}\, (\omega_{58a} + \omega_{67a},\, \omega_{57a} -
\omega_{68a}, \,\omega_{56a} + \omega_{78a})\,,
\end{eqnarray}
from the relation
\begin{eqnarray}
\label{tauA}
\sum_{A} \vec{\tau}^A \vec{A}^{A}_{m} &=&\sum_{s,t} M^{s t} \omega_{s t m}\,.
\end{eqnarray}
Taking into account Eq.~(\ref{deltaxsigma1})
% (Ref.~\cite{n2012scalars}, Eq.~(11)) 
one finds
\begin{eqnarray}
\label{tau1a}
\vec{\tau}^1\;\;\; \;&=& \vec{\tau}^{1\sigma} \, p_{\sigma} =
\vec{\tau}^{1\sigma}{}_{\tau} \,x^{\tau} \, p_{\sigma}\;, \nonumber\\
\vec{\tau}^2\;\;\;\; &=& \vec{\tau}^{2\sigma} \, p_{\sigma} =
\vec{\tau}^{2\sigma}{}_{\tau}\, x^{\tau} \; p_{\sigma}\,,
\nonumber\\
\vec{\tau}^{1\sigma}{}_{\tau}\,&=& \frac{1}{2}\, (e^{5}{}_{\tau} f^{\sigma 8} - 
e^{8}{}_{\tau} f^{\sigma 5} - e^{6}{}_{\tau} f^{\sigma 7} + e^{7}{}_{\tau} f^{\sigma 6},
\nonumber\\
\quad && e^{5}{}_{\tau} f^{\sigma 7} - 
e^{7}{}_{\tau} f^{\sigma 5} + e^{6}{}_{\tau} f^{\sigma 8} - e^{8}{}_{\tau} f^{\sigma 6},
\nonumber\\
\quad &&e^{5}{}_{\tau} f^{\sigma 6} - 
e^{6}{}_{\tau} f^{\sigma 5} - e^{7}{}_{\tau} f^{\sigma 8} + e^{8}{}_{\tau} f^{\sigma 7}),
\nonumber\\
\vec{\tau}^{2\sigma}{}_{\tau}&=& \frac{1}{2}\, (e^{5}{}_{\tau} f^{\sigma 8} - 
e^{8}{}_{\tau} f^{\sigma 5} + e^{6}{}_{\tau} f^{\sigma 7} - e^{7}{}_{\tau} f^{\sigma 6},
\nonumber\\
\quad && e^{5}{}_{\tau} f^{\sigma 7} - 
e^{7}{}_{\tau} f^{\sigma 5} - e^{6}{}_{\tau} f^{\sigma 8} + e^{8}{}_{\tau} f^{\sigma 6},
\nonumber\\
\quad &&e^{5}{}_{\tau} f^{\sigma 6} - 
e^{6}{}_{\tau} f^{\sigma 5} + e^{7}{}_{\tau} f^{\sigma 8} - e^{8}{}_{\tau} f^{\sigma 7})
\,.
\end{eqnarray}
The expressions for $f^{\sigma}{}_{m}$ are correspondingly 
\begin{eqnarray}
\label{fma} 
f^{\sigma}{}_{m}&=& \,(\vec{\tau}^{1\sigma}{}_{\tau}\, \vec{\cal{A}}^{1}_{m} +
\vec{\tau}^{2\sigma}{}_{\tau}\, \vec{{\cal A}}^{2}_{m})\, x^{\tau}\,.
\end{eqnarray}
Expressing the two $SU(2)$ gauge fields, $\vec{A}^{1}_{m}$ and $\vec{A}^{2}_{m}$, with 
$\omega_{s t m}$ as it is required in Eq.~(\ref{A1A2}), then using for each $\omega_{s t m}$ 
the expression presented in Eq.~(\ref{omegaabe1m}), in which  $f^{\sigma}{}_{m}$ is
replaced by the relation in Eq.~(\ref{fma}), then taking for $f^{\sigma}{}_{s} $ 
$= f \delta^{\sigma}_{s}$, where $f$ is a scalar function of the coordinates $x^{\sigma}$, 
$\sigma =(5,6,\dots,8)$ (in this case $e^{s}{}_{\mu} =$ 
$ -\delta^{m}_{\mu} e^{s}{}_{\sigma} f^{\sigma}{}_{m} $, Eq.~(\ref{fe1})), it follows 
after a longer but straightforward calculation that
\begin{eqnarray}
\label{testA}
\vec{A}^{1}_{m} &=&  \vec{\cal{A}}^{1}_{m}\,,\nonumber\\
\vec{A}^{2}_{m} &=&  \vec{\cal {A}}^{2}_{m}\,.
\end{eqnarray}
One obtains this result for any component of $A^{1i}_{m}$ and $A^{2i}_{m}$, $i=1,2,3$, 
separately.

It is not difficult to see that the gauge fields, which are superposition of $\omega_{st m}$, 
$(s,t)=(5,6,\dots,d)$, 
demonstrate in $d=(3+1)$ the  isometry of the space of $SO(d-4)$,
%where $SO((d-1) +1)$ of any  even $d$-dimensional space-time with the metric of
 Eq.~(\ref{fe}), with
%!!drop this away!
%
\begin{eqnarray}
e^{s}{}_{\sigma}=
f^{-1}\begin{pmatrix} 1 &0&0& \cdots 0 \\
0 & 1&0& \cdots 0\\
0&0&1&\cdots 0\\
&&&\cdots 0\\
&&&\cdots 0\\
0&0&\cdots&\;\;\;\;1 
\end{pmatrix}
\,, %\quad 
\nonumber\\ 
  \\
f^{\sigma}{}_{s}=
f \begin{pmatrix} 1 &0&0& \cdots 0 \\
0 & 1&0& \cdots 0\\
0&0&1&\cdots 0\\
&&&\cdots 0\\
&&&\cdots 0\\
0&0&\;\;\;\cdots&\;\;\;\;1 
\end{pmatrix}
\,.\nonumber
\label{fegen}
\end{eqnarray}
The space breaks into  $SO(3+1) \times SO(d-4)$
and $f $ is any scalar field of the coordinates: 
\begin{equation}
\label{maxsym}
f = f(\frac{\sum_{\sigma} (x^{\sigma})^2}{\rho_{0}^2})\,,
\end{equation}
while $\rho_{0}$ is the radius of the $(d-4)$ sphere and 
\begin{equation}
\label{fesigmam}
f^\sigma{}_m =  \sum_{A}\, \vec{A}^{A}_m \,\vec{\tau}^{A \sigma}{}_\tau \, x^\tau\,,
\end{equation}
where  $ \vec{A}^{A}_m$ are the superposition of $\omega^{st}{}_{m}$, 
$$ A^{Ai}_{m}=
\sum_{s,t} c^{Ai}{}_{st}\, \omega^{st}{}_{m},$$ demonstrating the symmetry of  
space with  $s\ge5$.
This illustrates the proof of the statement in the section~\ref{proof}.

\section{Relations between vielbeins and spin connections for scalars}
\label{scalars}

The {\it spin-charge-family} theory offers the explanation for the origin of the Higgs's scalar 
and the Yukawa couplings: The scalar gauge fields - the gauge fields of the charges described
by the two kinds of the Clifford algebra objects~\cite{norma2014MatterAntimatter,IARD2016},  
$\gamma^a$'s and $\tilde{\gamma}^a$'s,  (Eq.~(\ref{wholeaction})) - take care of masses 
of spinors after the electroweak break. 

We discuss here only the relation between vielbeins and spin connections for scalars the 
charges of which have the same origin as the charges of the vector gauge fields and only 
as long as  $(d-4)$ space manifests the isometry presented in 
Eqs.~(\ref{deltaxsigma}-\ref{ef}) with  the choice of $f^{\sigma}{}_{s} $ 
$= f \delta^{\sigma}_{s}$,  Eq.~(\ref{fespecial}), (which solves the Killing equation (\ref{Killing}), 
if $f$ is the scalar function of the coordinates $x^{\sigma}$). We do not include fermion 
sources which would change the symmetry of $(d-4)$ space, while $f^{\sigma}{}_{m} $ 
are (in the low energy regime) weak fields.
This section is only to point out the differences between the relation of spin connection - vielbeins
for vector and scalar gauge fields.

 Let us add that while the spin of the vector gauge fields in $(3+1)$ determines with respect to 
the space index $m=(0,1,2,3)$ the $SU(2)\times SU(2)$ 
structure of their spin, the space index $s$ of the superposition of the scalar spin connection 
fields -  $\sum_{t,t'} c^{Ai t t'}\,\omega_{t t' s}$ - manifests for $s=(7,8)$ the weak and hyper 
charges of the Higgs's scalar: ($\pm\frac{1}{2},\mp \frac{1}{2}$), respectively. Superposition
of the spin connection fields with the space indices $>8$ take care of transitions from the matter 
to antimatter and back, contributing to the matter-antimatter asymmetry of our universe.

To find the relation between vielbeins and spin connections we need to express the curvature 
$ R^{\sigma}{}_{\tau \sigma \tau'} g^{\tau \tau'}$ 
for  ($d-4$) space, where the Riemann tensor  and 
$\Gamma^{\sigma}{}_{\tau \sigma'}$ for this space are 
\begin{eqnarray} 
R^{\sigma}{}_{\tau \sigma' \tau'} &=&  \Gamma^{\sigma}{}_{ \tau [\tau', \sigma']}
+  \Gamma^{\sigma}{}_{\tau'' [\sigma'}\,  \Gamma^{\tau''}{}_{\tau \tau']},
\nonumber\\
 \Gamma^{\sigma}{}_{\tau \sigma'}& =&\frac{1}{2}\,  
g^{\sigma \tau'}\, (g_{\sigma' \tau'},_{\tau} + g_{\tau \tau'},_{\sigma'} - 
g_{\tau \sigma'},_{\tau'})\,,
\label{Rscalar}
\end{eqnarray}
 in terms of vielbeins $g^{\sigma \tau } =f^{\sigma}{}_{s} \, f^{\tau s}$, which is in our case
 $g^{\sigma \tau } =f^{2}\, \eta^{\sigma \tau}$, while $g_{\sigma \tau } =f^{-2} \,
\eta_{\sigma \tau}$ (${},_{\delta}$  again denotes the derivative with respect to 
$x^{\delta}$ and ${}_{[\; ]}$ the anti symmetrization with respect to particular two indexes)
and compare this expression with the corresponding one when $R$ is expressed with spin 
connections (and with the vielbeins).
\begin{eqnarray}
R              &=&  \frac{1}{2} \, \{ f^{\alpha [ a} f^{\beta b ]} \;(\omega_{a b \alpha, \beta} 
- \omega_{c a \alpha}\,\omega^{c}{}_{b \beta}) \} + h.c. \,.
\label{Romega}
\end{eqnarray}

One finds that $ \Gamma^{\sigma}{}_{\tau \sigma'}$ is for $f^{\sigma}{}_{s}= 
f\, \delta^{\sigma}_{s}$ equal to $ \Gamma^{\sigma}{}_{\tau \sigma'}$ $=f^{-1}\, 
(\delta^{\sigma}_{\sigma'}\, f,_{\tau} +  \delta^{\sigma}_{\tau}\, f,_{\sigma'} - 
 \eta_{\sigma' \tau}\, f^{,\sigma})$, while one finds for $\omega^{st}{}_{s'}= - (f^{,t}\,
\delta^{s}_{s'} - f^{,s}\, \delta^{t}_{s'})$ and for  $\omega^{st}{}_{\sigma}=
\omega^{st}{}_{s'} \,e^{s'}{}_{\sigma} =- f^{-1} (f^{,t}\,
\delta^{s}_{\sigma} - f^{,s}\, \delta^{t}_{\sigma})$.

It then follows for $R= R^{\sigma}{}_{\tau \sigma \tau'}\, g^{\tau \tau'}$, Eq.~(\ref{Rscalar}),
since  $\Gamma^{\sigma}{}_{ \tau [\tau', \sigma]}\,g^{\tau \tau'}= 2\, (d-4-1)\, ( f,_{\tau}
\,f^{,\tau} - f\, f,_{\tau}{}^{,\tau} )$ and
$ \Gamma^{\sigma}{}_{\tau'' [\sigma}\,  \Gamma^{\tau''}{}_{\tau \tau']}\,g^{\tau \tau'}=$
$( -1 +d-4) \,(2-d-4)\, f,_{\tau}\,f^{,\tau} $, that
\begin{eqnarray} 
R &=& R^{\sigma}{}_{\tau \sigma \tau'}g^{\tau \tau'}\nonumber\\
 &=&(d-4-1)\,\{ [2-(d-4-2)]\,\cdot f,_{\tau}
\,f^{,\tau}- 2\,\cdot f\, f,_{\tau}{}^{,\tau}\}\,.\nonumber\\
\label{Rscalarspecial}
\end{eqnarray}
We take into account Eq.~(\ref{omegaabe}) and evaluate  Eq.~(\ref{Romega}), obtaining   
\begin{eqnarray*}
 f^{\sigma [ s} f^{\tau t ]} \;\omega_{s t \sigma, \tau} &=& 
2\, (d-4-1)\, ( f,_{\tau} \,f^{,\tau} - f\, f,_{\tau}{}^{,\tau} ),\\%and
 f^{\sigma [ s} f^{\tau t ]} \;(-) \omega_{t' s \sigma}\,\omega^{t'}{}_{t \tau}&=&
( -1 +d-4) \,(2-d-4)\, f,_{\tau}\,f^{,\tau}], 
\end{eqnarray*}
what leads to 
\begin{multline}
 \frac{1}{2} \, \{ f^{\sigma [ s} f^{\tau t ]} \;(\omega_{s t \sigma, \tau} 
- \omega_{t' s \sigma}\,\omega^{t'}{}_{t \tau}) \} + h.c. =\\ 
(d-4-1)\,\{ [2-(d-4-2)]\,\cdot f,_{\tau}
\,f^{,\tau}- 2\cdot \,f\, f,_{\tau}^{,\tau}\}\,.
\label{nospinor}
\end{multline}

We conclude: If  $f^{\sigma}{}_{s}=
 \delta^{\sigma}_{s}\, f$, where $f = \,f(x^{\tau} x_{\tau})$, then 
both expressions for the curvature of $(d-4)$ space - the one with the metric tensor 
(Eq.~{\ref{Rscalar}}) and the one with the spin connection (Eq.~\ref{Romega})  - 
lead, as expected to the same expression
\begin{eqnarray} 
R&=&R^{\sigma}{}_{\tau \sigma \tau'}g^{\tau \tau'}\nonumber\\
 &=&
%2\,(d-4-1)\,\{ [2-(d-4-2)]\, f,_{tau}\,. f,^{\tau}- f\, f,_{\tau}{}^{,\tau}\} &=& 
\frac{1}{2} \, \{ f^{\sigma [ s} f^{\tau t ]} \;(\omega^{s t}{}_{ \tau, \sigma} 
+ \omega_{st' \sigma}\,\omega^{t'}{}_{t  \tau}) \} + h.c.\,,
\label{Rscalarspecial1}
\end{eqnarray}
where 
\begin{eqnarray} 
\omega^{st}{}_{\sigma}&=&
\omega^{st}{}_{s'} \,e^{s'}{}_{\sigma} = - f^{-1} (f^{,t}\,
\delta^{s}_{\sigma} - f^{,s}\, \delta^{t}_{\sigma})\,.
\label{Rscalarspecial2}
\end{eqnarray}
The result is valid also for the case that vielbeins and spin connections depend on the
coordinates of $(3+1)$ space: $f=f(\rho, x^m)$, $\omega_{s t t'}=\omega_{s t t'} 
(x^{\sigma}, x^m) $, $m=(0,1,2,3)$, $(s,t,t')= (5,6,..,d)$. 

That spin connections and vielbeins lead to the same Lagrange density in $(d-4)$ 
space, although as expected,  contributes to better understanding how in the low energy regime, 
after the electroweak break, scalar fields expressed with spin connections  $\omega_{stt'}$, 
$t'=(7,8)$,  offer the explanation for Higgs's scalar and the Yukawa couplings~\cite{IARD2016,JMP2015}.

\section{Conclusions}
\label{conclusion}

In the Kaluza-Klein theories the vector gauge fields - the gauge fields of the charges originating
in higher dimensional spaces - are represented through the vielbeins $f^{\sigma}{}_{m}$
(Eq.~(\ref{ef})) or rather with the corresponding metric tensors (Eqs.~(\ref{gmdown},%
\ref{gmup})). In the {\it spin-charge-family} theory the vector gauge fields are expressed 
as superposition of the spin connection fields $A^{Ai}_{m}$ $=\sum_{t,t'} c^{Ai t t'}$  
$\omega_{t t' m}$. This presentation offers an elegant and transparent understanding of the 
appearance of the vector gauge fields $A^{Ai}_{m}$, the charges of which originate in this
theory (and in the Kaluza-Klein theories) in higher dimensional spaces, while dynamics is 
determined in $(3+1)$.

Also the scalar (gauge) fields of the  {\it spin-charge-family} theory originate in higer dimensional 
spaces, offering the explanation for the origin of the Higgs's scalar and Yukawa couplings of
 the {\it standard model} - when the scalar gauge fields of both charges, $S^{st}$ and 
$\tilde{S}^{st}$ (Eq.~(\ref{twoclifford})), are taken into account~\cite{IARD2016}. Their 
dynamics is (like in
the case of the vector gauge fields) determined in $(3+1)$.
We discuss in this paper only gauge fields of $S^{st}$ for either vector or scalar 
fields.

We presented the proof, that the vielbeins $f^\sigma{}_m$ (Einstein index 
$\sigma \ge 5$, $m=0,1,2,3$)  lead in $d=(3+1)$ to the vector gauge fields, which are the 
superposition of the spin connection fields $\omega_{st m}$:  $f^\sigma{}_m=  \sum_{A}\,
 \vec{A}^{A}_m$ 
$\vec{\tau}^{A \sigma}{}_{\tau}\, x^{\tau}$, with $A^{Ai}_{m}=\sum_{s,t} c^{Ai}{}_{st}\,
 \omega^{st}{}_{m}$, when the metric in $(d-4)$, $g_{\sigma \tau}$, is invariant under the
coordinate transformations $x^{\sigma'} = x^{\sigma} + \sum_{A,i,s,t} \varepsilon^{A i}\,
(x^{m})\,c^{A i}{}_{st}$  $E^{\sigma s t} (x^{\tau})$ and 
$\sum_{s,t} c^{A i}{}_{st} \, E^{\sigma s t} = \tau^{A i \sigma}$, while  $\tau^{A i \sigma}$  
solves the Killing equation~(\ref{Killing}): 
$D_{\sigma}\, \tau^{A i}_{\tau} + D_{\tau} \tau^{A i}_{\sigma} =0\,$ ($D_{\sigma}\,
 \tau^{A i}_{\tau} = \partial_{\sigma}\, \tau^{A i}_{\tau} - \Gamma^{\tau'}_{\tau \sigma} 
\tau^{Ai}_{\tau'})$.

 We demonstrated for the case when $SO(7,1)$ breaks into $SO(3,1)\times SU(2)\times SU(2)$
that $\sum_{A,i} \tau^{Ai}\,  A^{Ai}_{m}= \sum_{s,t} S^{s t}\,\omega_{s t m}$ and that the 
effective action in flat $(3+1)$ space for the vector gauge fields is  $\int \,d^{4} x \, \{ - \frac{1}{4} \, 
F^{A i}{}_{m n} $ $F^{A i m n}\,\} $, where  $F^{Ai}{}_{mn} = \partial_{m} A ^{Ai}_{n}- 
\partial_{n} A ^{Ai}_{m} - i f^{Aijk} \, A^{Aj}_{m} $ $A^{Ak}_{n}$, and $f^{Aijk}$ are the 
structure constants of the corresponding gauge groups. 

The generalization of the break of $SO(13,1)$ into $SO(3,1)\times SU(2)\times SU(2) \times 
SU(3) \times U(1)$, used in the {\it spin-charge-family} theory, goes equivalently. In a general case 
one has $\sum_{A,i} \tau^{Ai}\, 
 A^{Ai}_{m}= \sum^{*}_{s,t} S^{s t}\,\omega_{s t m}$, where ${}^{*}$  means that the
summation concerns only those $(s,t)$, which appear in $\tau^{Ai}= \sum_{s,t} c^{Ai}{}_{st}
\,S^{st}$. These vector gauge fields $A^{Ai}_{m}$, expressible with the spin connection fields, 
$A^{Ai}_{m}= \sum_{s,t} c^{Ai st} \,\omega_{s t m}$, offer an elegant explanation for the
appearance of the vector gauge fields in the observed $(3+1)$ space.
The proof is true for any $f$ which is a scalar function of the coordinates $x^{\sigma},
 \sigma \ge 5$. 

We demonstrated also the relation between the spin connection fields and vielbeins for the scalar 
fields. While for the vector gauge fields the effective low energy action is in $d=(3+1)$ equal to 
$\int E' d^{4} x \, \{-\frac{1}{4}  F^{A i}{}_{m n} F^{A i m n}\} $ - where 
$F^{Ai}{}_{mn} = \partial_{m} A ^{Ai}_{n}- \partial_{n} A ^{Ai}_{m}
- i f^{Aijk} \, A^{Aj}_{m} \,A^{Ak}_{n}\,$, 
$A^{Ai}_{m}= \sum_{s, t} \,c^{Ai}{}_{st}\, \omega^{st}{}_{m}\,$, $E'=1$ in flat $(3+1)$ 
space, $\tau^{Ai} =  \sum_{s, t}\, c^{Ai}{}_{st} \,S^{st}$ and $f^{Aijk}$ are structure 
constants of the corresponding gauge groups  - it follows for the scalar fields  that, 
Eqs.~(\ref{Rscalar},\ref{Romega}),
\begin{eqnarray*} 
R&=& \{ \Gamma^{\sigma}{}_{ \tau [\tau', \sigma]}
+  \Gamma^{\sigma}{}_{\tau'' [\sigma}\,  \Gamma^{\tau''}{}_{\tau \tau']}\,\}\,g^{\tau \tau'}\nonumber\\
 &=&\frac{1}{2} \, \{ f^{\sigma [ s} f^{\tau t ]} \;(\omega^{s t}{}_{ \tau, \sigma} 
+ \omega_{st' \sigma}\,\omega^{t'}{}_{t  \tau}) \} + h.c.\,.
\end{eqnarray*}
  (The corresponding action is proportional  to $\int E\, d^{d-4} x\,$ $ R$).
Similar relation follows also for the superposition of the spin connection fields. 

If $\omega_{st' \sigma}$ depend on $x^{m}$ 
($x^{m}$ are coordinates in ($3+1$) space), the scalar fields are the dynamical fields in 
$(3+1)$, explaining, for example, after the break of the starting symmetry, the appearance 
of the Higgs's scalars and the Yukawa couplings~% Scalar fields contribute  as well to 
%the matter-antimatter asymmetry, of dark matter and others~
\cite{IARD2016,norma2014MatterAntimatter,JMP2015,JMP2013,n2012scalars}.

All these relations are valid as long as spinors and vector gauge fields  are weak 
fields in comparison with the fields which force $(d-4)$ space to be curled. When all these fields,
with the scalar gauge fields included,  start to be comparable with the fields (spinors or scalars), 
which determine the symmetry of $(d-4)$ space, the symmetry of the whole space changes. 

%We indeed assume an almost (Ref.~\cite{HNnjp2011}) compacitified space, which means that 
%there are sources which force $(d-4)$ space to compactify. It is shown in 
%Ref.~\cite{hn2006,HNnjp2011} as an example that for $f=(1+ \frac{\rho^2}{(2\rho_0)^2})$ the 
%symmetry of the space with the coordinate $x^{\sigma},\sigma =(5),(6)$, is a surface $S^2$ with 
%one point missing. 

%The proof offers indeed no 
%surprise due to the fact that the spin connection fields $\omega_{abc}$ are expressible with the 
%vielbeins as presented in (Eq.~(\ref{omegaabe}). 

% 17.03.2017 at 11:45
% Think about equations of motion of the scalar fields and also what the tilde scalar fields 
% are.
%\end{document}
\appendix

\section{Derivation of the equality $\vec{A}^{1}_{m} =  \vec{\cal{A}}^{1}_{m}$}
\label{Ader}

We demonstrate for the case $A^{11}_m$ $=(\omega_{58m} - 
\omega_{67m})$,  Eq.~(\ref{A1A2}), that this  $A^{11}_m$ is equal  to 
${\cal A}^{11}_{m}$, appearing in  Eq.~(\ref{fma})
\begin{equation}
f^\sigma{}_m = \sum_{A,i}\, \mathcal{A}^{Ai}_m \tau^{Ai\sigma}{}_\tau x^\tau\,.
\label{fmsigma}
\end{equation}

When using Eq.~(\ref{omegaabe1m}) for $A^{11}_{m}=\omega_{58m} - \omega_{67m}$
we end up with the expression
\begin{eqnarray}
A^{11}_m &=& %\frac{i^2}{2}
 \frac{1}{2E} \biggl\{ 
  f^\sigma{}_m [e^8{}_\sigma \partial_\tau (Ef^{\tau 5}) -
  e^5{}_\sigma \partial_\tau(Ef^{\tau 8})]\nonumber\\  
 && \qquad  - f^\sigma{}_m [e^7{}_\sigma \partial_\tau (Ef^{\tau 6}) -
  e^6{}_\sigma \partial_\tau(Ef^{\tau 7})] \nonumber\\
 &&\qquad  + e^5{}_\sigma \partial_\tau [E(f^\sigma{}_m  f^{\tau 8} -
    f^\tau{}_mf^{\sigma8})]  \nonumber\\
  &&\qquad - e^6{}_\sigma \partial_\tau [E(f^\sigma{}_m  f^{\tau 7} -
    f^\tau{}_mf^{\sigma7} )]  \nonumber\\
 &&\qquad  - e^8{}_\sigma \partial_\tau [E(f^\sigma{}_m  f^{\tau 5}  -
    f^\tau{}_mf^{\sigma5} )]\nonumber\\ 
 &&\qquad  + e^7{}_\sigma \partial_\tau [E(f^\sigma{}_m  f^{\tau 6}) -
    f^\tau{}_mf^{\sigma6}] \biggr\}. 
\label{derivation1}
\end{eqnarray}
Inserting for $f^\sigma{}_m$ the expression from  Eq.~(\ref{fmsigma}) we obtain,
when taking into account Eq.~(\ref{tau1a}),
\begin{eqnarray}
\label{A11}
A^{11}_m &=& 
%-\frac{1}{2} \frac{1}{2E} \sum_i \mathcal{A}^{11}_m \biggl\{ 
%  \tau^{1i\sigma}{}_{\tau'} x^{\tau'} [e^8{}_\sigma \partial_\tau (Ef^{\tau 5}) -
%  e^5{}_\sigma \partial_\tau(EF^{\tau 8})
%  - e^7{}_\sigma \partial_\tau (Ef^{\tau 6}) +
%  e^6{}_\sigma \partial_\tau(EF^{\tau 7})] \nonumber\\
% &&\qquad  + e^5{}_\sigma \delta^{\tau'}_\tau E 
 % (f^{\tau 8} \tau^{1i\sigma}{}_{\tau'} - f^{\sigma8}
 % \tau^{1i\sigma}{}_{\tau'})
%+ e^5{}_\sigma x^{\tau'} \partial_\tau [E(f^{\tau 8}  \tau^{1i\sigma}{}_{\tau'}
%-  f^{\sigma8} \tau^{1i\sigma}{}_{\tau'})] \nonumber\\
% &&\qquad-e^6{}_\sigma  E 
%  (f^{\tau 7} \tau^{1i\sigma}{}_{\tau'} - f^{\sigma7}
%  \tau^{1i\sigma}{}_{\tau'})
%- e^6{}_\sigma x^{\tau'} \partial_\tau [E(f^{\tau 7}  \tau^{1i\sigma}{}_{\tau'}
%-  f^{\sigma7} \tau^{1i\sigma}{}_{\tau'})] \nonumber\\
% && \qquad-e^8{}_\sigma  E 
%  (f^{\tau 5} \tau^{1i\sigma}{}_{\tau'} - f^{\sigma5}
%  \tau^{1i\sigma}{}_{\tau'})
%- e^8{}_\sigma x^{\tau'} \partial_\tau [E(f^{\tau 5}  \tau^{1i\sigma}{}_{\tau'}
%-  f^{\sigma5} \tau^{1i\sigma}{}_{\tau'})] \nonumber\\
% && \qquad+e^7{}_\sigma  E 
%  (f^{\tau 6} \tau^{1i\sigma}{}_{\tau'} - f^{\sigma6}
%  \tau^{1i\sigma}{}_{\tau'})
%+ e^7{}_\sigma x^{\tau'} \partial_\tau [E(f^{\tau 6}  \tau^{1i\sigma}{}_{\tau'}
%-  f^{\sigma6} \tau^{1i\sigma}{}_{\tau'})] 
%\biggr\}\,. 
 \partial_8 (f_{5m})  - \partial_5 (f_{8m})  - \partial_7 (f_{6m}) + \partial_6 (f_{7m}) \,.
\end{eqnarray}
Inserting Eq.~(\ref{fmsigma}), in which we take into account Eq.~(\ref{tau1a}) 
as well as  that $e^{s}{}_{\sigma} = f^{-1} \delta^{s}_{\sigma} $ and 
$f^{\sigma}{}_{s} = f \delta_{s}^{\sigma} $,  into  
Eq.~(\ref{A11}), % after using  Eq.~(\ref{tau1a}) in  which we
   we and up with   
\begin{equation}
A^{11}{}_m =  \sum_{A,i} \mathcal{A}^{Ai}{}_m\, \delta^{A}_{1} \,\delta^{i}_{1}\,.
\label{A11proof}
\end{equation}
%
%where $ \mathcal{C}^{1i}$ can be read off Eq.~(\ref{A11}).
%Taking into account in Eqs.~(\ref{tau1a},\ref{A11}) that $f^{\sigma}{}_{s}= f \delta^{\sigma}_{s}$
%and $e^{s}{}_{\sigma}= f^{-1} \delta^{s}_{\sigma}$ we find that most of terms in $\mathcal{C}^{11}$
%cancel each other. The only term, which remains, originates in terms from coordinate derivatives, 
%leading to
%
Similarly one obtains for the gauge fields of both subgroups $SU(2)\times SU(2)$ of the group
$SO(4)$
\begin{eqnarray}
A^{Ai}{}_m =  \sum_{B,j} \mathcal{A}^{Bj}{}_m \,\delta^{A}_{B}\, \delta^{i}_{j}\,.
\label{ASO(4)proof}
\end{eqnarray}
Similar derivations go for any $SO(n)$.

%(!!! Do the SO(6) and the break of $SU(2)\times SU(2)$ to $SU(2)\times U(1)$!!!)

%
%Correct notation!
%while we found that
%$\mathcal{C}^{12}= 0 = \mathcal{C}^{13}$. 
%
%Recognizing that $\mathcal{C}^{2i}$ contribute to  $A^{11{}}_{m}$ nothing, we can conclude 
%that
%$A^{11}{}_m =  \mathcal{A}^{11}{}_m$.

%One easily see that to the expressions for $A^{Ai}{}_{m}$ only $\mathcal{C}^{Ai}$ contribute, while
%all $\mathcal{C}^{Bj}$, $B \ne A$ and $j \ne i$ contribute nothing. 
%This completes the proof that $\vec{A}^{A}{}_{m} = \mathcal{\vec{A}}^{A}{}_{m}$, for all the 
%gauge fields $\vec{A}^{A}{}_{m}$ of the charges $\vec{\tau}^{A}$, Eq.~(\ref{tau1a}).

%
 
%
\end{document}